\documentstyle[aps,psfig]{revtex}

\newcommand{\patisalam}{SU(2)_L\times SU(2)_R \times SU(4)_C}

\begin{document}

\title{SUPERSYMMETRIC UNIFICATION AT THE MILLENIUM\footnote
{Invited Talk presented at WHEPP-6 Workshop on High Energy
Physics, Chennai, January 3-15, 2000}}

\author{Charanjit S. Aulakh}

\address{Department of Physics, Panjab University,\\
Chandigarh, INDIA\\E-mail: aulakh@panjabuniv.chd.nic.in}


\maketitle
\begin{abstract}
\tightenlines
 We argue that  the discovery of
neutrino mass effects at Super-Kamiokande implies a clear
logical chain leading from the Standard Model, through the MSSM
and  the recently developed Minimal Left Right Supersymmetric
models  with a renormalizable
see-saw mechanism for neutrino mass, to Left Right symmetric SUSY GUTS :
in particular, $SO(10)$ and $SU(2)_L \times SU(2)_R\times SU(4)_c$.
The progress in constructing such GUTS explicitly is reviewed and their
testability/falsifiability by lepton flavour violation and 
proton decay measurements emphasized. Susy violations of the
survival principle and the interplay between third generation Yukawa coupling
unification and the structurally stable IR attractive features
of the RG flow in SUSY GUTS are also discussed .
\end{abstract}

\section{Introduction}
\tightenlines
This workshop is proceeding on the morning of the new millenium
in the shadow of the towering cranes of a mammoth construction
project. Thus it is entirely appropriate to review the progress
in  the ongoing `petaproject' of the
``palazzo GUT'' whose patroness (the santissima susy) has after 20
years of fervid courtship/worship still not lost her charisma
for the high energy theory community. Although several
``sky-high'' colonne of the Susy-GUT palazzo are now firmly in
place (Force and third generation yukawa unification, Matter
unification and the neutrino mass unification mass connection
(nozze dei neutrini)) {\it{il palazzo manca tetto e muri }}
(the palace still lacks a roof and walls). Thus it is still open
to the gales of speculation habitual in its clime. On the
other hand the discovery of neutrino mass effects at
SuperKamiokande in the 100 milli eV range have made LR
supersymmetric models and their GUT relatives hot favourites for
a direct look through the windows of the palazzo. Finally
progress has been made in the last years of the millenium in
identifying the characteristic doors through which we may enter the
palazzo of Susy unification: namely the generic signals provided
by lepton flavour violation and proton decay . The questions of
fermion masses of the first and second generations , fermion
mixings etc are the walls and services of the palace whose
emplacement is an open project for the new millenium.

In a recently published review talk I have covered the
connections between the MSSM with R parity and LR supersymmetric
unification in prolix detail\cite{dae}. On the other hand that
review  devoted little attention to the progress in
understanding the gross features of the fermion mass spectrum
and the generic lepton flavour violating signals for susy
unification. Therefore the main points of the previous talk are
here related telgraphically and the freed space used to discuss  fresh
topics within the space limitations imposed. My notations, abbreviations
and conventions are those of that article\cite{dae} and will not be
repeated here.

Let us begin with the implications of the Super-Kamiokande
discovery of beyond SM effects\cite{superk}. 
In the SM neutrino mass can be introduced only via the 
non-renormalizable operator $(H^{\dagger} L)^2/M$. The presence of
the scale M indicates new physics beyond the SM. 
Super-Kamiokande data interpreted as evidence of tau
neutrino mass $\sim 10^{-1.5} eV$ implies 
the scale of the new physics leading to neutrino mass is  
$M\sim 10^{14\pm .5}$ GeV . Such a large scale can be plausibly
separated from the Electroweak scale by the introduction of
Supersymmetry which can prevent quadratic corrections ($\sim M^2$) to light
boson masses . Thus one may argue that the heirarchy problem is
no longer the hypothetical one posed by GUTs but is now posed by the
SM and the Super-Kamiokande data !

Since it offers an appealing and flexible escape route from the
heirarchy problem, Susy has enjoyed a 20 year old vogue that
shows no sign of abating.  
There are good reaons for this fascination . As reviewed in
detail in \cite{dae} it was  predicted (in
1982 !) \cite{marsen,einjn1}  that
only susy unification would be compatible with the data if it
turned out that the top quark mass and Weinberg angle were as
large as they were ultimately found to be (i.e $m_t\sim 200
$GeV and $Sin^2 \theta_W\sim .23$). This been strikingly
vindicated by the precise EW data accumulated by LEP I and II.
Furthermore the larger value of the unificication scale in the
Susy case explains why baryon decay has not long been seen.
Successful and unique gauge unification by minimal SUSY GUTs 
constitutes the first pillar of the ``Palazzo Susy'' and was the
reason for the refocussing of interest on GUTs from the
beginning of the nineties\cite{amlflo}.

The unfication of gauge couplings naturally begs for an explanation
via spontaneous symmetry breaking of a larger GUT gauge symmetry
in which the Electroweak-QCD physics is embedded. 
The canonical and minimal possibilities for the GUT gauge
symmetry are the Pati-Salam symmetry 
$G_{224}=G_{PS}=SU(2)_L \times SU(2)_R\times SU(4)_c$ and SU(5)
and also the rank 5 group $Spin(10)$ ($SO(10)$ with fermions)
in which both can be embedded. $Spin(10)$ is also minimal in
that it has a simple gauge group which
unifies the fermions of the SM, together with the right handed
neutrino needed to explain neutrino mass effects convincingly,
in a single irrep. It can accomodate several different breaking
chains by use of appropriate Higgs multiplets and shall be the
favoured model of this review.

The singular properties of the neutrino ( it is the only neutral
fermion of the SM , it is bereft of a chiral partner, and its 
minimal mass operator - the $d=5$ operator above - violates the
anomaly free global $B-L$ symmetry of the SM ), the
indications of its mass  found by
Super-Kamiokande as well as the explanation of the solar
neutrino deficit via neutrino oscillations all obtain a natural
and elegant explanation in terms of a `see-saw' between the
neutrino and its righthanded partner\cite{gmann} . The extra field $\nu_R$
is naturally and necessarily
present in LR symmetric GUTs where the maximal parity violation
of the SM is viewed as arising from spontaneous breaking of a
parallel $SU(2)_R$ symmetry present at high energies. The `fatso'
scale of the seesaw is moreover precisely the scale $M_{B-L}$ at
which the {\it{gauged}} B-L symmetry of LR models breaks to the
weakly violated global symmetry of the SM. Explicitly , the
presence of $\nu^c_l \sim \nu_R^*$ allows one to write the mass
term and yukawa interaction (absent in the SM ) as 

\begin{equation}
(M_{\nu^c})_{ij} \nu^c_{Li} \nu^c_{Lj} + h_{ij} H^{\dagger} L_i \nu^c_{Lj}
\label{mterms}
\end{equation}
Integrating out the heavy Neutrino gives the neutrino mass
operator 

\begin{equation}
-h_{ij} (M_{\nu^c}^{-1})_{jk} h_{kl} H^{\dagger} L_i H^{\dagger} L_l
\label{numasopdet}
\end{equation}
This is the type I see-saw\cite{gmann}. In case the left neutrinos
acquire (small!) Majorana masses $M_{\nu_L}$ from some other
source then they will add to the above contribution giving the
so called Type II see saw mechanism \cite{rabigoran}.
The mass $M_{\nu^c}$ may itself be taken to arise via
spontaneous symmetry breaking of the $B-L$ symmetry via vevs of
$SU(2)_R\times U(1)_{B-L}$ multiplets : which must be triplets if
the interaction is to be renormalizable, but may be taken to be
doublets if one permits higher dimensional operators to do the
job. 

{\bf{R-Parity and LR symmetry :}} 
The presence of scalar partners of the fermions of the SM in
the MSSM allows renormalizable interactions which
violate the crucial $B-L$ symmetry which protects the current
universe from evaporation into radiation. These violations are
thus strongly constrained by our persistence. The simplest way
to remove them is by noting that in their absence the MSSM
enjoys the discrete mutiplicative symmetry $R_p=(-)^{3(B-L) +
2S}$\cite{rabirpar} .
Imposing this symmetry thus furnishes a natural if, {\it
{ad hoc}}, justification for the B-L symmetry of the SM. The form
of the symmetry now leads to an unexpected brownie point for LR
theories : $R_p$ is effectively a part of their gauge symmetry .
Moreover if one implements the see-saw using a renormalizable term
to generate the $\nu^c$ mass then the required Higgs fields
$\Delta^c$  are $SU(2)_R$ triplets with $B-L=- 2$ and 
{\it{their vevs cannot violate R parity}}.  Explicitly, the
couplings are 

\begin{equation}
h_{ij}^a L_i\phi_a L^c_j + f_{ij} L_i \Delta L_j + f^c_{ij}
L^c_i \Delta^c L^c_j +h.c
\label{yukawas}
\end{equation}

These give a large Majorana mass to the $\nu^c_L$
fields when the $\Delta^c$ field develops a large vev, while the
vev of the $\phi(2,2,0,1)$ field which is dominantly responsible for EW SSB
gives rise to a Dirac mass term between $\nu_L$ and $\nu^c_L$.
Thus a seesaw mechanism occurs very naturally as a
consequence of the hierarchy between $SU(2)_L\times U(1)_Y$ and 
$SU(2)_R\times U(1)_{B-L}$ breaking scales. Since the scalar potential in
general allows couplings of the form 

\begin{equation}
V= M^2 \Delta^2 + \Delta \phi^2 \Delta^c + ......
\label{type2}
\end{equation}

It follows that once $\phi,\Delta^c$ acquire vevs $\sim M_W,M$
respectively $\Delta$ acquires a vev due to the linear term generated :
\begin{equation}
<\Delta> \sim M_W^2/M
\label{deltvev}
\end{equation}

so that the seesaw is in general of Type II.

\section{Minimal Susy LR Models}

A significant advance \cite{abs} has been the realization that
when -as is generically the case and as is now experimentally
indicated - the scale of $B-L$ violation is $M_{B-L}>>M_W$,
phenomenological constraints and the structure of the SUSY
vacuum ensure that R-parity is preserved. 
Viable Minimal LR Supersymmetric Models (MSLRM) have been
constructed in detail \cite{abs,ams,amrs}
 and embedded in GUTS while retaining these
appealing properties.

  Since the argument for R-parity exactness
is so simple and general we present it first in isolation before
going on to the details of the MSLRMs. Given $M_{B-L} >> M_W\sim
M_S$ (the scale of SUSY breaking) it immediately follows that
the scalar partners of the $\nu^c_L$ fields also have positive
mass squares $\sim M_{B-L}^2 $ and hence are protected from
getting any vevs : modulo effects suppressed by these large
masses. Thus when the $\nu^c_L$ superfield is integrated out the
effective theory is the MSSM with R-parity and with $B-L$
violated only by (the SUSY version of) the highly suppressed
$d=5$ neutrino mass operator. As a result if , for any reason,
the scalar ${\tilde\nu}_L$ were to obtain a vev \cite{am82} the low energy
theory would contain an almost massless scalar (i.e a  ``doublet''
(pseudo) Majoron) in its spectrum. However  such a
pseudo-Majoron is  conclusively ruled out by LEP.
 These arguments based on decoupling are very
robust, see  \cite{dae,abs,ams,amrs} for details.
 To sum up, quite generally  :

 {\it{The low energy effective theory of MSLRMs with a
renormalizable is the MSSM with exact R parity so 
the Lightest Supersymmetric Particle (LSP) is stable }}. 

The detailed structure of MSLRMs has been established in the
series of papers cited above for the generic case (now favored
strongly by experiment) when the scale of Right handed (B-L
breaking) physics is high. We conclude that
the fields of generic MSLRMs 
(in addition to the supersymmetrized anomaly free set of fields
of the LR model $(Q,Q^c,L,L^c,\phi,\Delta,{\bar{\Delta}},{\Delta}^c
{\overline{\Delta}}^c)$ ) are one of the following :
 
{a)} Introduce a pair $\Omega(3,1,0,1)\oplus
\Omega^c(1,3,0,1)$ 0f $SU(2)_{L/R}$ triplet fields . Then one can
achieve SSB of the LR symmetry via the $\Omega$s and separately
the SSB of the $B-L$ symmetry, at an independent scale $M_{B-L}$
 by the $\Delta^c,{\bar\Delta}^c$ fields.

{b)} Stay with the minimal set of fields , but (reasoning
that small non-renormalizable corrections must be counted when
the leading effects are degenerate) include the next order $d=4$
operators allowed by gauge invariance in the superpotential.
These operators are of course suppressed by some large scale M
and may be thought to arise either from Planck scale physics
$(M\sim M_{Planck})$ or when one integrates out heavy fields in
some GUT in which the LR model is embedded ($M\sim M_X)$. The 
principal effect of allowing such terms is that the charge
breaking flat direction is lifted and one obtains a
phenomenologically viable low energy effective theory (with
characteristic additional fields at the scale $M_{B-L}^2/M_R$ :
see below). 

{c)} Finally one may introduce a parity odd singlet in
either of cases a) b) which we shall for convenience refer to as
cases $a')$ and $b')$. This case is not  academic or non-minimal since
such parity odd singlets(POS) arise very naturally when one embeds
these models in $SO(10)$.

The symmetry breaking in these models has been analyzed in
considerable detail using the help of the theorem \cite{luty} which labels
the vacuum manifold of SUSY theories by the chiral gauge
invariants left unfixed by the vanishing if the F terms on the
vacuum manifold. One then finds that the above parity preserving
scenario is convincingly realised .

{\bf{Survival Principle Violations :}} The detailed analysis of
symmetry breaking also allows one to calculate the mass spectrum
of the theory\cite{abmrs,dae}.
 Besides, the usual particles of the SM and their
superpartners at $M_S$ one finds  that certain
superfields associated with $SU(2)_R\times U(1)_{B-L}$ breaking
remain relatively light and for favourable values of the
parameters may even be detectable at current or planned
accelerators. In cases $a)$ and $a')$ one finds that a complete
supermultiplet with the quantum numbers of $\Omega(3,1,0,1)$
has a mass $M_{B-L}^2/M_R$ . If $M_{B-L}<<M_R$ then these
particles could be detectable.  However given the expectation of
$M_{B-L} > 10^{14} GeV$ from neutrino mass this does not appear
to be a likely possibility. In Cases $b)$ and $b')$ (which one
may consider as the truly minimal alternative one) finds instead
that the entire slew of fields
$\Delta,{\bar{\Delta}},\delta^c_{--},{\bar{\delta} }^c_{++}, 
\delta^c_{0}+ {\bar{\delta} }^c_{0}, H_u', H_d'$ have masses
$\sim M_R^2/M$. If , for instance, $M\sim 10^{19} GeV$ and
$M_R\sim 10^{11} GeV$ then these particles could conceivably be
detectable specially because they include exotic particles
with charge 2 which are coupled to the usual light fermions of
the model.In the above $H_u',H_d'$ are a pair Higgs doublets left
over after the fine-tuning to keep one pair of doublets
light out of the four (i.e two bidoublets) 
introduced to allow sufficient freedom in the tree level Yukawa couplings .

The {\it{reason}} for the lightness (noticed from the beginning \cite{amso10})
 of the exotic super
multiplets has been under appreciated in the past. It is due to
a generic feature of SUSY lagrangians where gauge invariance and
the constraint of
renormalizability of  the superpotential 
often leads to an absence of cubic
couplings for some fields.

This important feature of symmetry breaking and mass spectra in
Susy theories has recently been emphasized in \cite{surviv}.
Normally the masses of the non goldstone parts of a Higgs
multiplet get masses of the same magnitude as the symmetry
breaking vev.  Thus it is considered reasonable when undertaking
RG analyses to assume that the mass spectrum obeys this 
``survival principle '' and compute the RG flow even
without calculating the mass spectrum explicitly. However it was
noticed long ago\cite{amso10} that there tend to be large supermultiplets
which remain light after high scale susy breaking involving vevs
for their GUT partners. The reason for this can be clarified by
simple example. Consider a Wess-Zumino model with a single
chiral superfield in which, for some reason,  the superpotential
contains a quadratic and higher than cubic terms(suppressed by
 powers of some large mass $M>> m$) but no cubic term. 

\begin{equation}
W = m\Phi^2  + \sum_{n>0}{\phi^{n+3} \over M^n}
\end{equation}

When $\Phi$ gets a vev ($\sim  \sqrt{mM}$) the effective cubic coupling is thus
$\sim ({{<\Phi>}\over M})^{n_{min}} <<1$ so that the mass of the
residual non goldstone super multiplet is $<< \quad <\Phi>$. 

For a gauge example consider a U(1) model with two fields $\phi,
{\bar \phi}$ with opposite charges . Then the superpotential
takes the form

\begin{equation}
W = m \phi{\bar \phi} - {{(\phi{\bar \phi})^2}\over {2M}}
\end{equation}

 This then imples that the vevs of both fields are $ \sqrt{mM}$
while the effective cubic coupling $\sim \sqrt{m/M}$ is very small
if the scale $M >> m$ as is natural to suppose for non
renormalizable couplings . Indeed one finds that while one
multiplet($\Phi - {\bar \Phi}$) is the super Higgs and has a mass
$g <\phi>\sim g\sqrt{mM} $ the other $\Phi + {\bar \Phi}$ has a much smaller
mass $\sim m$. These considerations apply directly in Susy LR
models since the $\Delta$ multiplets (in SO(10) the 126's) have
no cubic couplings . This leads to light doubly charged exotics
and makes it abundantly clear that RG analyses based on a blind
application of the survival principle are can be erroneous.
These observations can obviously have critical implications for low
energy SUSY phenomenology.

\section{LR Susy Guts}

As we have seen, LR SUSY models are natural candidates for SUSY
unification which accommodates neutrino mass. Thus it is natural
to consider further unification in which the various factors of
the LR symmetry group are unified with each other. The two most 
appealing possibilities are unification within the Pati-Salam
group $\patisalam $ and $SO(10)$. The multiplets
{\bf{45,210}} of $SO(10)$ contain parity odd singlets
\cite{mohaparid} and the Pati-Salam gauge group is a subgroup of
$SO(10)$. Thus the study of $SO(10)$ unification 
teaches one much about the Pati-Salam
case as well. Therefore we\cite{abmrs,dae} have re-examined  SO(10) SUSY
unification  keeping in view the progress in understanding of LR
SUSY models detailed above and developed a minimal
SO(10) Theory of R-Parity and Neutrino mass with the
appealing features of automatic R-parity conservation . A
detailed and explicit study of the SSB at the GUT scale and
various possible intermediate scales was performed. The  mass
spectra in various cases could be  explicitly computed. In
particular the light supermultiplets with possibly
low intermediate scale masses ($\sim M_R^2/M_{PS}, M_{PS}^2/M_X$
etc.) that often arise in SUSY GUTS (see \cite{amso10} for an early
example involving SO(10)) were determined. With these computed
(rather than assumed spectra) a preliminary one-loop RG survey
of coupling constant unification in such models was carried out.
We find that an appealing and viable model may be constructed 
using the Higgs multiplets ${\bf{45(A),54(S),126(\Sigma),{\overline
{126}}({\overline{\Sigma}}),10(H)}}$.  The reader is referred to
\cite{dae,abmrs} for details.

\section{Fermion Masses, Yukawa Unification and IR attractors}

The striking and consistent unification of gauge couplings in
the MSSM naturally raises the
question of the behaviour of the other parameters of the
MSSM at high energies. Even in the SM there are already $\sim
15$ other parameters. Allowing for neutrino masses and mixings
and for Susy breaking parameters raises this number to $\sim
10^2$. Thus a separation of parameters into different classes
corresponding to the gross and fine structure of the model is
necessary to make progress.
After gauge couplings and vector boson masses the most important
parameters are the fermion masses.  These are conveniently
divided into three subproblems :

{a)} Third generation masses/Yukawa couplings : Since the
third generation masses , in particular $m_{top} $ are so much
larger than the the others it is plausible that they are
connected with the gross structure of the fermion mass matrix
and of symmetry breaking.

{b)} The second and first generation masses and the mixing
matrix in the quark sector obey : 
$m_{u,d,c,s.e.\mu} << m_{t,b,\tau},V_{CKM} <<1 $.
and may plausibly be considered to arise as a next order effect
due to radiative or other suppressed corrections. Recently some
progress has been made along these lines in models closely
related to the ones favoured here \cite{bdmrad}.

{c)} As discussed above neutrino masses and
mixings are an independent issue due to the neutrino's peculiar
properties. Moreover they inevitably imply physics beyond the SM
and hence fall in an independent category. The question of how
the large neutrino mixing angles apparently favoured by data could be
compatible with GUTs and the small mixings in the quark sector
has attracted much model building attention recently. However
these discussions lie out of the minimalistic gaze of this review.

In the MSSM there is an important new parameter freedom
necessarily present: the ratio of the vevs of the two Higgs
doublets $tan \beta = v_u/v_d$. This freedom breaks the strict
correlation between the yukawa couplings $h_{t,b,\tau}$ and the
corresponding masses $m_{t,b,\tau}$.  The minimal versions
of the basic types of GUT models imply strict relations between
yukawa couplings at the unification scale and the consistency of
these relations with the low energy data is then an important
constraint on these models.

In the case of the Pati-Salam model with fermion masses
$(2,2,4)_F (2,2,{\bar 4})_F <(2,2,1>$ 
arising from bidoublet vevs and also in minimal SO(10) models
with fermion masses $16_F 16_F <10_H>$ the relations
$h_t=h_b=h_{\tau}$ follow naturally. SU(5) invariant mass terms
$10_t 10_{t^c} <{\bar{5_{\bar H}}}> + 10_{b,\tau} 5_{\tau^c,b^c}<5_H>$
however imply only $h_b=h_{\tau}$.

The implications of the so called Pendleton-Ross \cite{pross}
fixed point, the Hill\cite{hill} effective fixed point, and
their manifold generalizations, for
our understanding of the labyrinthine question of fermion masses
are so profound, and shed so much light on these
vexed questions that I will review them in some detail. My
treatment is largely based on the excellent review of Schrempp
and  Wimmer\cite{schwim}.

Consider first the evolution of the couplings $h_t,g_3$ ,
ignoring other couplings . It is convenient to form the ratio
$\rho_t= h_t^2/g_3^2$ and to trade the evolution parameter
$\mu$ for the asymptotically free coupling $g_3$ with which it
is monotonically correlated. The evolution equation at one loop is

\begin{equation}
{d\rho_t \over {dln g_3^2}} = -2 \rho_t (\rho_t - {7 \over 18})
\end{equation}

Clearly there is a fixed point at $\rho_{tf} =7/18$ and is easily
seen to correspond to a top quark mass of $126 Sin \beta $ GeV.

Furthermore, it can be shown \cite{hill,schwim} that for
sufficiently large initial value $\rho_0$
  that the low energy value of $\rho$ obeys the Hill bound 

\begin{equation}
\rho < {7 \over {18}} (1 - ({\alpha_3^0 \over {\alpha_3}})^{7/9})^{-1} 
\end{equation}

This bound functions as an effective attractor for
the IR flow which slows down markedly in its vicinity  .

Next one can introduce the coupling $h_b$ and write the
corresponding RG equations

\begin{eqnarray}
{{d\rho_t} \over {dln g_3^2}} &=& -2 \rho_t (\rho_t +{\rho_b\over 6}
 - {7 \over {18}})\nonumber \\
{{d\rho_b} \over {dln g_3^2}} &=& -2 \rho_b (\rho_b +{\rho_t\over 6} 
  - {7 \over 18}) 
\end{eqnarray}

Thus one now has the following structural features :

$\bullet$ Infrared Attractive Fixed Point {\bf (0)}: $\rho_t=
\rho_b = 1/3$. 

$\bullet$ Strongly Attractive Infrared Attractive Fixed Line
{\bf (1)}: see Fig. 1 .

$\bullet$ Less Attractive IR Fixed Line {\bf (2)}: $\rho_t=
\rho_b $ (intersects {\bf{1}} in {\bf{0}}.

$\bullet$ Hill effective fixed line {\bf (0)}: outer arc in Fig.
1 .

Note that except {\it{on}} the fixed line 
 {\bf{2}} the flow is first attracted by
the fixed line {\bf{1}} and then along it to the fixed point. 

When one switches on the other gauge couplings $g_1,g_2$ the
fixed point {\bf{0}} generalizes to $\rho_1=\rho_2=\rho_{\tau}=0,
\rho_t=\rho_b=1/3$ while the structures {\bf{0,1,2,3}} become
embedded in an IR attractive surface onto which the flow
proceeds before going onto {\bf 1} and then along it into {\bf
0} unless initially (and then always ) on {\bf 2}. 
The role of the attractive fixed structures in inducing insensitivity
to the initial values of the couplings at high
energies is dramatically illustrated by considering a range of
$\rho_{t0},\rho_{b0}\in [0,25]$ i.e initial values at $ \mu_0=
e^{t_0} \sim 10^{16} $GeV . This
entire range of intial values flows into
$\rho_{t},\rho_{b}\in [0,1]$ at $\mu\sim 10^2 $GeV. Indeed
most of the parameter range (say $[.5,25]$) flows into a narrow
band between the fixed line {\bf{1}} and the  Hill quasi-attractor {\bf{3}}.
One can use the trivial relation 
$tan \beta = {m_t\over m_b} {\sqrt {\rho_b\over \rho_t}}$
to trade the fixed lines {\bf{1,3}} for lines in the
$tan\beta$-$m_t$ plane to obtain Fig. 2.
We know that almost any initial values will flow into the
narrow band between the lines {\bf{1,3}}. Thus this figure offers
us a structurally stable insight into the structure of the
parameter space of the MSSM. It is worth mentioning
that the effect of including the electroweak gauge couplings is
to narrow this band and make the Hill line {\bf 3} more
attractive \cite{schwim}. From Fig. 2 it is clear that there
are two narrow ranges of $tan \beta$ ($tan\beta \in [1,4]$ or
$tan\beta \in [42-66]$) compatible with
experimental value of $m_t\sim 175 GeV$. These ranges have a strong
dependence on $m_b^{expt}$ but only a weak dependence on
$m_s,\alpha_s$. Fig. 3 illustrates various possibilities.


The role of the fixed lines {\bf 1,3} in permitting the viability
of GUT initial conditions like $h_{\tau}^0= h_b^0$ can be
clarified by considering the RG equation for the  ratio 
$R=h_{\tau}/ h_b$ . 

\begin{equation}
{dR\over {dt}} = {R\over {16\pi^2}} ( h_t^2 -{16\over 3} g_3^2 +
3 (h_b^2 -h_{\tau}^2) + {4\over 3} g_1^2)
\label{RRG}
\end{equation}

This first order DE has {\it two} conditions
on it set by the GUT condition $R^0=1$ and the experimental
constraint $R(m_t) =m_b/m_{\tau}$. To satisfy them the value of
$h_t^0$ (the free parameter in (\ref{RRG})) must be fine tuned .
On the other hand $h_t$ has its own RGE and final condition to
satisfy. The role of the IR attractive structures is precisely
to make the flow insensitive to the value of $h_t^0$ provided it
is greater than some minimum value . 
If one further imposes the equality of $h_b^0=h_{\tau}^0= h_t^0$
as required by minimal SO(10) one finds that one flows into the
close vicinity of the IR fixed point {\bf 0} and $tan \beta \sim
50$. {\it{So that Susy SO(10) models prefer the high $tan \beta$
solution}}.

The state of the art in the study of the RG structure of Susy
GUT models can be glimpsed from the results of a recent
paper\cite{tata} which also analyses the compatibility of the
above discussion with radiative electroweak symmetry breaking (EWRSB).
Recall that if the top Yukawa coupling $h_t$ is large enough then
$m_{H_u}^2$ flows to smaller values and can turn negative
leading to EWRSB\cite{agaume}.  The breaking is
constrained to obey $m_{H_d}^2 -m_{H_u}^2 > M_Z^2$.
However when one attempts to combine this  scenario ,
considered by many to be one of the most attractive bonuses of
the MSSM, with the picture of the structurally stable attractor
governing the third generation fermion masses one runs into an
obstacle. It has been shown \cite{borsoleprok} that minimal susy
SO(10) with universal soft susy breaking terms suffers from 
problems in obtaining simultaneous consistency with experimental bounds
the branching ratio for $b\rightarrow s\gamma$ and the
cosmological bound $\Omega _{LSP} h^2< 10^{-1}$ since the 
LSP is essentially the Bino. Also  $m_{H_d}^2 <m_{H_u}^2 $ so that there
is no EWRSB. However it was soon shown \cite{murolepro} that non-universalty
in the soft breaking terms introduced naturally via the small
D term vev $<D_X> \sim M_S$
when SO(10) breaks to the MSSM (here X=-2Y + 5(B-L) is the
diagonal generator broken while reducing the rank of the gauge
group from 5 to 4) was sufficient to split the initial values of
the soft masses for $H_u,H_d,{\tilde Q}, {\tilde U} , {\tilde D} $
etc so that the constraints of the model with universal soft
terms were evaded and EWRSB was achievable. 

The recent work of \cite{tata} has shown that this program can
indeed be inplemented successfully over a considerable region of
the soft susy parameter space. The authors of \cite{tata}
carried out a scan of the soft susy parameter space in the
minimal SO(10) model i.e they generated random initial values
for $m_{1/2}, M_{16}, m_{10}, m_D, A_0$ (the soft trilinear
coupling) and integrated the two loop RG equations for 26
couplings and masses . For an appreciable range of parameters
they found that  i) the electroweak gauge
couplings unify exactly and the strong coupling is within $10\%$,
ii) $h_{t,b,\tau}$ unify within $5\%$ yielding a value of $tan
\beta 48= \pm 4$, iii) The LSP is mostly a Higgsino so that the
constraints from the $b\rightarrow s\gamma$ and $\Omega_{LSP}$
are obeyed, iv) EWRSB is achieved. A representative plot of one
of their successful models is shown as Fig. 4. The successful
gauge and yukawa unification is obvious as is the non
universality in the soft masses at the high scale and the EWRSB
at low energies. The outlook for these models is thus very
favorable since they combine so many desirable features and yet
run the gauntlet of experimental and cosmological constraints
successfully. 

Further recent work on generation of realistic fermion
masses and mixings for the first and second generations via
radiative corrections is very promising and
appealing{\cite{bdmrad}}. However a detailed scan of the
parameter space at the level of the above analysis of
\cite{tata} remains to be done.

\section{Lepton Flavour Violating Signals for Susy Guts}

A notable development in the basic picture of Susy Guts in the
nineties has been the realization \cite{hkr} that in
contrast to non SUSY theories SUSY GUTS inevitably imply ver
exotic  lepton flavour violating signals suppressed only by
 inverse powers of the low
scale $M_S$ and by CKM mixing angles . 

The basic argument is very simple. Firstly in the SM the
$B,L_{e,\mu,\tau}$ symmetries are `accidental ' consequences of the
gauge symmetries . When the SM is Grand Unified the fact that
GUT gauge symmetries rotate quarks into leptons imples that
Baryon number and the three lepton numbers are violated .
However since they are symmetries of the SM the effects of the
violation are represented by effective operators suppressed by
inverse powers $M_{GUT}$.

On the other hand , in the MSSM these quantum numbers are no
longer automatic and it is necessary to introduce R parity 
($R_p=(-)^{3(B-L) + 2S}$!) to
forbit the violation of B and $L_{total}$ by $d=4$ operators.
{\it{ Howeve,r since $L=L_{e} + L_{\mu}+L_{\tau}$,
 this symmetry does not ensure that the individual
Lepton flavours are conserved !}}. Therefore, in Susy GUTs the
presence of lepton flavour violation in the soft susy breaking sector
will inevitably leak into the fermion interaction . Now since
this effect must vanish as $M_S\rightarrow \infty$ it follows that
it is suppressed only by inverse powers of $M_S$ ! Thus, for
example, in minimal Susy SU(5) one finds that the branching ratios for
the exotic processes $\mu\rightarrow e\gamma, \tau \rightarrow
\mu\gamma$ are within an order of magnitude of the current
experimental limits due to mixing between the different right
handed sleptons. As one considers more general non
renormalizable superpotentials with couplings
between GUT Higgs and three fermion representations suppressed
by the Planck mass (in order to fit the light quark masses\cite{mura}) one
finds that the left handed sleptons also mix and lead to further
exotic processes such as $\mu\rightarrow e\gamma$ and electric dipole
moments for the neutron. For large values of $tan \beta$ the
rates are further enhanced. Furthemore, when one considers
models with a $\nu^c$ - as one must to make sense of neutrino
masses - left slepton mixing contributions again arise and raise
the value of exotic branching ratios such as 
$\mu\rightarrow e\gamma, \tau \rightarrow \mu\gamma$.  

The importance of these {\it generic} signals of susy unfication
can hardly be overemphasized. Indeed the thrust of this talk has
precisely been that the ``zeroth order'' version of Lepton
flavor violation, namely the effects of neutrino mass , calls
naturally for Susy. It is then gratifying to note that
conversely Susy unification can signal its presence clearly
through generic lepton flavour violating effects whose
decoupling scale is $M_S$ itself and that these signals are
already on the threshold of verifiability/falsifiability with
current dectectors. The reader is referred to \cite{dae} for 
the situation regarding
proton decay, regarding which we have nothing fresh to relate.

\section{Conclusions}

To summarize : besides the well established and appreciated
unification of representations and gauge couplings achieved by
SUSY GUTs , one also now sees that : 

$\bullet $ There is a clear logical chain leading from the SM with
neutrino mass to the Minimal Supersymmetric LR models with
renormalizable seesaw mechanisms developed in detail recently.

$\bullet $ These MSLRMs have the MSSM with R parity and  seesaw
neutrino masses and have quasi exact B-L as their effective low
energy theory. They can also have light Higgs
triplet supermultiplets in their low energy spectra, due to
violations of the survival principle in Susy theories, leading to
very distinctive experimental signatures.

$\bullet $ They can be embedded in GUTS based on the PS group or
SO(10). The former case may be more suitable in stringy
scenarios which so far disfavor light (on stringy scales) SO(10)
GUT Higgs of dimension $>54$. 

$\bullet $ The SSB in the SO(10) SUSY GUT has been worked out
explicitly and the mass spectra calculated. This allowed us to
perform a RG analysis based on calculated spectra leading to the
conclusion that $M_X\geq 10^{15.5} GeV$ while $M_R,M_{PS}\geq
10^{13} GeV$. With $M_X$ at its lower limit the $d=6$ operators
for nucleon decay can become competitive with the $d=5$
operators raisining the possibility that observation of
$p\rightarrow \pi^{0}e^+$ need not rule out SUSY GUTS after all.

$\bullet $ Fermion mass spectra can be compatible with charged
fermion mass data and neutrino mass values suggested by neutrino
oscillation data from super Kamiokande and Solar neutrino
oscillation experiments .

$\bullet $ Dimension five operators in theories with seesaw lead to
a remarkable connection between neutrino masses and nucleon
decay which constrains these models fairly tightly and makes
them testable by upcoming nucleon stability measurements.

$\bullet $ Further work on the doublet triplet splitting problem,
the question of fermion mass spectra , two loop RG analysis etc. 
is required.

$\bullet $ {\it{ Thus LR SUSY seesaw models and their GUT
generalizations look good. The show has just begun but it aint
over till the fat neutrina sings !}}

\eject
\centerline{\bf{Figure captions}}
\vskip .5 true cm

Fig. 1 Fixed lines in the $\rho_t-\rho_b$ plane.
The outer arc is  is the Hill effective fixed line ( {\bf{3}}). 
\vskip .5 true cm
Fig 2. : The lines {\bf{1}}(heavy line) {\bf{3}} (outer line)
translated to the $\tan\beta - m_t$ plane. After ref[21].
\vskip .5 true cm

Fig 3. Contours of constant $m_b$ in the $\tan\beta - m_t$
plane subject to tau-bottom unfication. Read $h_{b,\tau,t}$ for 
$\lambda_{b,\tau,t}$ (after Barger, Berger and Ohmann see Ref.[21]).
\vskip .5 true cm
Fig. 4. : RG flow in a minimal $SO(10)$ model. After Ref.[22]
 \vskip .5 true cm

\eject
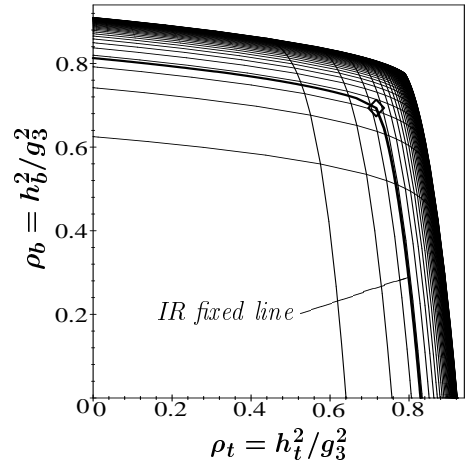
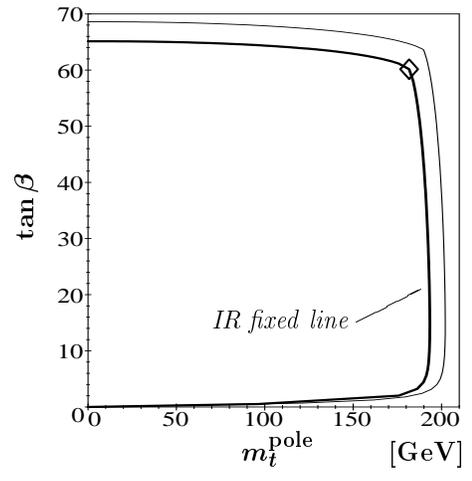
\end{document}